\documentclass[conference]{IEEEtran}

\IEEEoverridecommandlockouts

\usepackage{cite}
\usepackage{amsmath,amssymb,amsfonts}
\usepackage{algorithmic}
\usepackage{graphicx}
\usepackage{textcomp}
\usepackage{xcolor}
\usepackage{booktabs}
\usepackage{array}
\usepackage{url}
\usepackage{tikz}
\usetikzlibrary{arrows.meta,positioning,fit,calc,shapes.geometric}
\def\BibTeX{{\rm B\kern-.05em{\sc i\kern-.025em b}\kern-.08em
    T\kern-.1667em\lower.7ex\hbox{E}\kern-.125emX}}
\newcommand{\lumosfull}{Language-Model Unified Machine-Readable Operating-System Semantics}

\begin{document}
\bstctlcite{IEEEtran:BSTcontrol}

\title{LUMOS: A Semantic Operating-System Layer for Accessibility-Grounded AI Agents}

\author{
\IEEEauthorblockN{Yogeswar Reddy Thota}
\IEEEauthorblockA{\textit{Department of Electrical and Computer Engineering}\\
\textit{University of Texas at Dallas}\\
Richardson, TX, USA\\
YogeswarReddy.Thota@UTDallas.edu}
}

\maketitle

\begin{abstract}
Current operating systems expose interfaces optimized for human users but not
for AI agents. Humans benefit from pixels, icons, windows, visual grouping,
mouse movement, and keyboard shortcuts; AI agents instead need compact
semantic state, grounded actions, and reliable feedback. As a result, many
computer-use agents are forced to interpret screenshots, OCR output, and
visual crops, introducing high token costs, visual ambiguity, latency, and
coordinate uncertainty. This paper introduces LUMOS (\lumosfull), a semantic
interaction layer between AI agents and operating systems. LUMOS converts native
accessibility metadata and browser UI structures into machine-readable
semantic blueprints with stable identifiers, roles, names, values, bounds,
and action affordances. It also supports live semantic pointer grounding by
querying the UI element under or near the cursor through operating-system
automation APIs. An LLM then acts through an accessibility-grounded
observe--act loop using constrained visible-UI primitives rather than
application-specific scripts. LUMOS does not claim to replace visual agents;
instead, it reduces dependence on screenshots when operating systems already
provide semantic structure. These results suggest a path toward AI-native
operating systems and machine-readable interaction layers.
\end{abstract}


\section{Introduction}
Artificial intelligence has rapidly entered software engineering, search,
communication, writing, and knowledge work. Many of these deployments occur
inside environments that are already text-first or API-first: command-line
interfaces, programming languages, web search boxes, chat interfaces, and
structured documents. In these settings, the model can often act directly
on symbolic text. The same is not true for the general desktop.

For decades, mainstream operating systems such as Windows, macOS, and Linux
desktop environments have optimized user interfaces for humans. The term
``user interface'' itself reflects this design center. A human benefits from
visual grouping, color, depth, animation, iconography, spacing, and hover
states. An AI agent, however, does not need a blue button to be blue in
order to infer that it is clickable. It needs the button's purpose, state,
location, and permissible actions. A screenshot may contain this information,
but it embeds it in a high-entropy visual representation.

Recent computer-use agents commonly rely on screenshots and visual-language
models to identify controls and infer actions. This approach is powerful
because it can operate on arbitrary visual surfaces. It is also expensive
and brittle: the agent must parse pixels, infer UI semantics, estimate
coordinates, and ignore visual decoration. Existing benchmarks show that
open-ended computer tasks remain difficult for state-of-the-art agents,
especially when tasks span desktop applications, operating-system state,
and long-horizon workflows \cite{osworld,webarena,windowsworld}.

This paper explores a different intuition: operating systems already expose
machine-readable descriptions of many visible interfaces. Accessibility
and automation frameworks were originally developed for screen readers,
assistive technologies, and UI testing. On Windows, Microsoft UI Automation
(UIA) exposes desktop UI elements as structured objects arranged in a tree,
with properties such as name, role, control type, value, bounding rectangle,
and supported patterns \cite{microsoft_uia_overview,microsoft_uia_tree}.
Browsers similarly expose the Document Object Model (DOM) and accessibility
trees. These structures are closer to what an AI agent needs than a raw
screenshot.

We propose LUMOS, a semantic blueprint layer for LLM-driven operating-system
interaction. LUMOS does not attempt to replace the OS kernel or bypass the
visible interface. Instead, it adds a machine-facing interaction plane over
human-first software. The layer observes the current native or web UI,
extracts a compact blueprint, assigns stable element identifiers, asks the
LLM for a single valid action, executes that action through visible UI
mechanisms, and observes again. The result is an observe--plan--act loop in
which the LLM remains responsible for strategy, while LUMOS provides
grounding, safety, memory, and execution.

\textbf{Contributions.} This paper makes four contributions:
\begin{itemize}
\item \textbf{Semantic operating-system layer:} a middleware that
    transforms native operating-system accessibility metadata into
    machine-readable semantic blueprints for AI agents.
    \item \textbf{Live semantic pointer grounding:} cursor-position-aware
    semantic interaction using live UI Automation queries, including
    ElementFromPoint-style grounding of the interface under the pointer.
    \item \textbf{Accessibility-grounded observe--act loops:} an alternative
    to screenshot/OCR-centric computer use in which the agent navigates and
    controls software through semantic roles, values, bounds, and structured
    actions.
    \item \textbf{Toward AI-native computing:} evidence that existing
    operating systems can expose a machine-readable interaction plane for AI
    agents without requiring applications to be redesigned.
\end{itemize}

\section{Motivation: Human-First OS, Machine-First Agents}
Human-centered GUI design hides system complexity behind visual metaphors.
Files become icons, applications become windows, and operations become
buttons, menus, and gestures. This abstraction is excellent for people, but
it is not necessarily optimal for AI agents.

An LLM agent must answer questions such as: What application is active? What
controls exist? Which control accepts text? What text is already present?
Which action is safe? Has the previous action completed the task? A screenshot
contains visual evidence for many of these questions, but not in a directly
symbolic form. The model must spend tokens and compute on perception that the
operating system may already know.

For example, when a cursor is over a button, operating-system and application
frameworks often know the element bounds, accessible name, role, state, and
supported interaction patterns. UIA exposes much of this information to
automation clients. A semantic blueprint can therefore represent a window as
the following compact structure:

\begin{verbatim}
A2: role=Document
    name="Text editor"
    value="hello"
    bounds=(120, 180, 900, 600)
\end{verbatim}

This is more directly useful to an LLM than a screenshot crop of Notepad. The
model can choose:

\begin{verbatim}
{"action":"type_text","target_id":"A2",
 "text":"Hello from LUMOS"}
\end{verbatim}

The system then performs the grounded interaction. The sweet spot is not
giving the LLM unrestricted access to the machine. It is giving the LLM a
safe, structured, visible, and reversible interface to what a human could
see and do.

\begin{figure}[t]
\centering
\resizebox{\columnwidth}{!}{%
\begin{tikzpicture}[
    font=\scriptsize,
    plane/.style={draw, rounded corners=3pt, thick, minimum width=4.6cm,
        minimum height=1.05cm, align=center},
    bridge/.style={draw, rounded corners=3pt, thick, fill=green!10,
        minimum width=4.6cm, minimum height=0.82cm, align=center},
    arrow/.style={-{Stealth[length=2mm]}, thick},
    feedback/.style={-{Stealth[length=2mm]}, thick}
]
\node[plane, fill=blue!6] (human) at (0,2.15)
{\textbf{Human UI plane}\\pixels, windows, controls, mouse, keyboard};

\node[bridge] (lumos) at (0,0.55)
{\textbf{LUMOS semantic layer}\\UIA / DOM metadata $\rightarrow$ stable element IDs};

\node[plane, fill=orange!10] (agent) at (0,-1.05)
{\textbf{Agent interface plane}\\roles, names, values, bounds, actions};

\draw[arrow] (human.south) -- node[right, align=left]{extract\\semantics} (lumos.north);
\draw[arrow] (lumos.south) -- node[right, align=left]{compact\\blueprint} (agent.north);
\draw[feedback] (agent.west) .. controls (-2.6,-1.05) and (-2.6,2.15) ..
    node[left, align=center]{validated\\visible action} (human.west);
\end{tikzpicture}}
\caption{Conceptual motivation. LUMOS converts a human-facing visual interface
plane into a machine-readable agent interface plane without requiring the
application to be redesigned.}
\label{fig:motivation}
\end{figure}
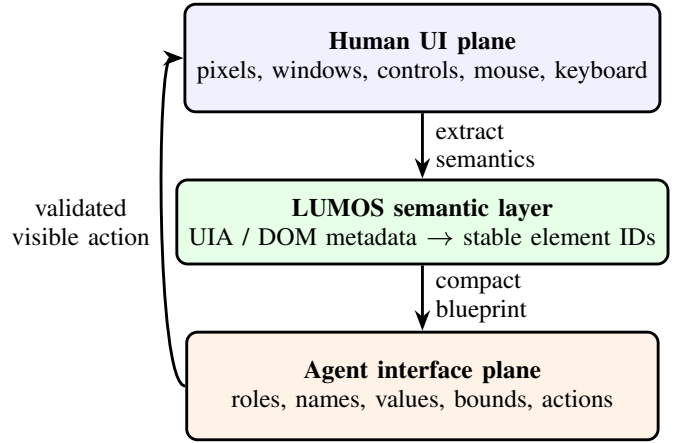

\section{From User Interfaces to Agent Interfaces}
The history of computing interfaces can be read as a sequence of abstractions
over machine operation. Command-line interfaces made computation accessible
through symbolic commands. Graphical user interfaces made computation
accessible through windows, icons, menus, and pointing devices. Touch
interfaces reduced the distance between perception and action, while voice
interfaces allowed users to express intent conversationally. Each stage
expanded who could use computers by changing the interaction plane.

We argue that AI agents motivate the next stage: \emph{agent interfaces}.
An agent interface is not a replacement for the human interface. Rather, it
is a parallel machine-readable plane through which an AI system can perceive
available state, understand actionable structure, and request constrained
operations. Future operating systems may therefore expose two coordinated
planes:

\begin{itemize}
    \item \textbf{Human interface plane:} windows, buttons, pixels, colors,
    layout, mouse movement, keyboard input, touch, and voice.
    \item \textbf{Agent interface plane:} semantics, accessibility trees,
    structured state, element roles, values, action affordances, safety
    policies, and machine-readable completion feedback.
\end{itemize}

LUMOS is an early prototype of the second plane. It does not require
applications to be rewritten for AI agents; instead, it reuses the semantic
metadata already exposed by operating systems and browsers. This positions
LUMOS as a machine-native interaction layer for future AI-native computing
environments.

\section{Background}
\subsection{UI Automation and Accessibility Trees}
Microsoft UI Automation is an accessibility framework that provides
programmatic access to most desktop UI elements and allows assistive
technologies and test scripts to inspect and manipulate interfaces
\cite{microsoft_uia_entry,microsoft_uia_overview}. UIA represents elements
in a tree rooted at the desktop, where application windows contain controls
such as menus, buttons, edit fields, lists, and documents
\cite{microsoft_uia_tree}. Each element can expose properties and control
patterns describing its semantics and behavior.

The important observation for AI agents is that UIA was not designed for LLMs,
yet it already approximates a symbolic interface layer. It can tell an agent
that a visible region is a text box, button, list item, document, or menu, and
can provide names, values, focusability, and coordinates. Similar principles
hold for web interfaces, where DOM nodes and browser accessibility trees
describe the functional structure of a page.

\subsection{LLM Agents and Computer Use}
LLM agents combine reasoning and action selection. ReAct demonstrated the
value of interleaving reasoning traces and actions in language-model-driven
decision making \cite{react}. Computer-use benchmarks such as WebArena and
OSWorld, and Mind2Web highlight the difficulty of grounding language
instructions in interactive web and desktop environments
\cite{webarena,osworld,mind2web}. Recent work continues to show that
cross-application desktop workflows are especially challenging
\cite{windowsworld}.

Many systems approach grounding through screenshots and visual-language
models. Operator, Claude computer use, OmniParser, ScreenAI, UI-TARS, and
Agent-S all demonstrate the importance of allowing agents to operate in
human-facing interfaces \cite{openai_operator,anthropic_computer_use,
omniparser,screenai,uitars,agent_s}. LUMOS is complementary but differently
positioned: it uses existing UI structure where available and reserves visual
methods for cases where semantic APIs fail. This can reduce context size,
improve action grounding, and make failures easier to debug. AutoGen and
related multi-agent frameworks provide orchestration patterns for LLM systems
\cite{autogen}; LUMOS instead focuses on the operating-system interaction
plane through which such agents can perceive and act.

Traditional robotic process automation is also related because it automates
desktop work, but it usually depends on predesigned workflows or brittle UI
scripts \cite{rpa_survey}. LUMOS targets a different layer: semantic
observation and visible action primitives that an LLM can use dynamically.

\begin{table*}[t]
\centering
\caption{Positioning LUMOS relative to vision-grounded computer-use agents.}
\label{tab:related}
\begingroup
\scriptsize
\setlength{\tabcolsep}{2pt}
\renewcommand{\arraystretch}{1.12}
\begin{tabular}{>{\raggedright\arraybackslash}p{0.13\linewidth}
                >{\raggedright\arraybackslash}p{0.14\linewidth}
                >{\raggedright\arraybackslash}p{0.14\linewidth}
                >{\raggedright\arraybackslash}p{0.13\linewidth}
                >{\raggedright\arraybackslash}p{0.09\linewidth}
                >{\raggedright\arraybackslash}p{0.10\linewidth}
                >{\raggedright\arraybackslash}p{0.14\linewidth}}
\toprule
\textbf{System} & \textbf{Examples} & \textbf{Interaction} & \textbf{Semantic Access} & \textbf{OCR} & \textbf{Pixels} & \textbf{Cursor Grounding}\\
\midrule
Vision-grounded agents & Operator, Claude computer use, UI-TARS & Screenshot/browser or desktop control & Inferred from pixels or model captions & Often useful or implicit & High & Typically visual/coordinate-based\\
Screen parsers & OmniParser, ScreenAI & Screenshot parsing and UI element detection & Predicted from visual regions & Often yes & High & Indirect through detected regions\\
Agent orchestration & AutoGen, Agent-S & Tool and agent coordination & Depends on tool environment & Depends on environment & Varies & Not the primary contribution\\
Semantic-grounded agents & LUMOS & UIA/DOM/accessibility blueprints and visible actions & Native roles, values, bounds, state & No, when semantic APIs are available & Low by default & Direct ElementFromPoint-style semantic query\\
\bottomrule
\end{tabular}
\endgroup
\end{table*}

\section{LUMOS Architecture}
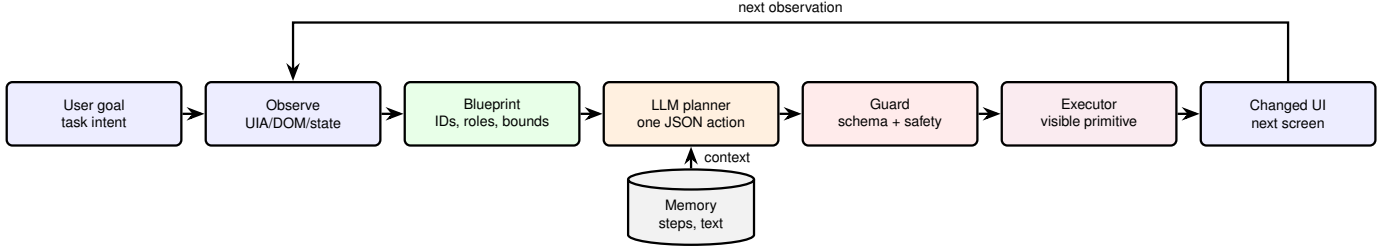
\begin{figure*}[t]
\centering
\resizebox{\textwidth}{!}{%
\begin{tikzpicture}[
    font=\sffamily\tiny,
    box/.style={draw, rounded corners=2pt, thick, minimum height=0.78cm,
        minimum width=2.15cm, align=center},
    obs/.style={box, fill=blue!7},
    sem/.style={box, fill=green!9},
    llm/.style={box, fill=orange!12},
    safe/.style={box, fill=red!8},
    act/.style={box, fill=purple!8},
    store/.style={draw, cylinder, shape border rotate=90, aspect=0.24,
        thick, fill=gray!10, minimum height=0.9cm, minimum width=1.55cm,
        align=center},
    arrow/.style={-{Stealth[length=2.2mm]}, thick},
    feedback/.style={-{Stealth[length=2.2mm]}, thick}
]

\node[obs] (goal) at (0,0) {User goal\\task intent};
\node[obs] (observe) at (2.45,0) {Observe\\UIA/DOM/state};
\node[sem] (blueprint) at (4.9,0) {Blueprint\\IDs, roles, bounds};
\node[llm] (planner) at (7.35,0) {LLM planner\\one JSON action};
\node[safe] (safety) at (9.8,0) {Guard\\schema + safety};
\node[act] (exec) at (12.25,0) {Executor\\visible primitive};
\node[obs] (ui) at (14.7,0) {Changed UI\\next screen};

\node[store] (memory) at (7.35,-1.25) {Memory\\steps, text};

\draw[arrow] (goal) -- (observe);
\draw[arrow] (observe) -- (blueprint);
\draw[arrow] (blueprint) -- (planner);
\draw[arrow] (planner) -- (safety);
\draw[arrow] (safety) -- (exec);
\draw[arrow] (exec) -- (ui);
\draw[arrow] (memory.north) -- node[right, xshift=1pt]{context} (planner.south);
\draw[feedback] (ui.north) -- ++(0,0.72) -| node[pos=0.25, above]{next observation} (observe.north);

\end{tikzpicture}}
\caption{End-to-end LUMOS data-flow pipeline. A user command is converted
into semantic observations from UIA, DOM/accessibility trees, system state,
and live pointer grounding. These are compacted into a semantic blueprint and
ID map, combined with task memory, given to the LLM planner, validated through
schema and safety checks, executed on the visible UI, and fed back into the
next observation.}
\label{fig:architecture}
\end{figure*}

LUMOS is organized as a layered observe--plan--act system, shown in
Fig.~\ref{fig:architecture}. In its default mode, the runtime observes
semantic UI state, exposes universal visible-action primitives, and asks the
LLM to choose the next step rather than dispatching a prewritten workflow.

\subsection{Perception Layer}
The perception layer reads the current system state and active interface.
For native Windows applications, it queries foreground windows and UIA trees.
For web pages, it queries the browser session and extracts DOM/accessibility
information. The result is normalized into a blueprint with fields such as:

\begin{itemize}
    \item element identifier, e.g., \texttt{A2} for native or \texttt{W3} for web;
    \item role or control type, e.g., button, document, edit, menu item;
    \item accessible name and current value;
    \item bounding rectangle or screen coordinates;
    \item window title, URL, and focus context;
    \item optional semantic hints relevant to the current goal.
\end{itemize}

The blueprint is intentionally compact. It excludes decorative visual details
unless they are needed for action. This makes it cheaper to send to an LLM
than a screenshot and easier to validate.

\subsection{Live Semantic Pointer Grounding}
Human users continuously combine vision with pointer position: moving the
cursor over an interface reveals what object is being targeted, whether it is
clickable, and what action may follow. LUMOS approximates this capability
semantically. Instead of cropping the screen around the cursor and asking a
vision model to infer meaning, the runtime can query the operating system for
the UI Automation element at or near a screen coordinate. In Windows terms,
this corresponds to ElementFromPoint-style grounding: a coordinate is mapped
to a semantic element with a role, name, state, value provider, and bounding
rectangle.

This mechanism makes the pointer itself part of the semantic interaction
plane. A model or runtime can ask not only ``what pixels are under the
cursor?'' but ``what UI element is under the cursor, what does it mean, and
what operations does it expose?'' The result is a live bridge between physical
input coordinates and machine-readable interface semantics. This is especially
useful when full-window blueprints are large, when an element is ambiguous,
or when the agent needs to confirm that a planned click is grounded in the
intended control.

\begin{figure}[t]
\centering
\resizebox{\columnwidth}{!}{%
\begin{tikzpicture}[
    font=\sffamily\scriptsize,
    box/.style={draw, rounded corners=2pt, thick, minimum width=2.25cm,
        minimum height=0.62cm, align=center},
    pix/.style={box, fill=red!7},
    sem/.style={box, fill=green!9},
    act/.style={box, fill=orange!12},
    arrow/.style={-{Stealth[length=2mm]}, thick}
]
\node[box, fill=blue!7] (xy) at (0,0) {Pointer coordinate\\$(x,y)$};

\node[pix] (crop) at (3.0,0.75) {Screenshot crop};
\node[pix] (ocr) at (5.9,0.75) {OCR / vision\\inference};
\node[pix] (guess) at (8.8,0.75) {Guessed target\\label + coordinate};

\node[sem] (uia) at (3.0,-0.75) {UIA hit test\\ElementFromPoint};
\node[sem] (element) at (5.9,-0.75) {Semantic element\\role, name, value};
\node[act] (target) at (8.8,-0.75) {Grounded ID\\\texttt{A7}/\texttt{W4}};

\draw[arrow] (xy.east) -- ++(0.45,0) |- (crop.west);
\draw[arrow] (crop) -- (ocr);
\draw[arrow] (ocr) -- (guess);
\draw[arrow] (xy.east) -- ++(0.45,0) |- (uia.west);
\draw[arrow] (uia) -- (element);
\draw[arrow] (element) -- (target);

\node[font=\sffamily\bfseries\scriptsize, text=red!70!black] at (5.9,1.45) {vision-first path: infer semantics from pixels};
\node[font=\sffamily\bfseries\scriptsize, text=green!45!black] at (5.9,-1.45) {LUMOS path: query semantics directly from OS};
\end{tikzpicture}}
\caption{Live semantic pointer grounding. LUMOS maps a physical pointer
coordinate to the UI Automation element under the cursor, avoiding a
screenshot-crop-and-OCR step when accessibility semantics are available.}
\label{fig:pointer-grounding}
\end{figure}
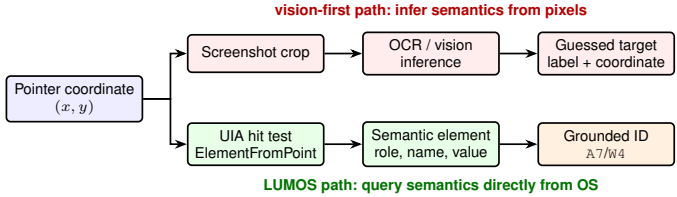

\subsection{Planner Layer}
The planner is an LLM prompted with the user goal, recent memory, and the
current blueprint. It must emit a single JSON action from a constrained schema.
This one-step discipline is important. It avoids hallucinated long scripts
and forces the agent to re-observe after each action, similar to how a human
checks the screen after clicking or typing.

\subsection{Universal Action Schema}
LUMOS exposes a small set of visible UI primitives:

\begin{itemize}
    \item \texttt{observe}: refresh perception;
    \item \texttt{open\_windows\_search}: open the visible Windows search overlay;
    \item \texttt{open\_app}: launch known safe applications;
    \item \texttt{click}, \texttt{double\_click}, \texttt{drag}: interact with
    identified elements;
    \item \texttt{type\_text}: type into a focused or targeted control;
    \item \texttt{set\_text}: replace existing text rather than append;
    \item \texttt{press\_key}: submit or navigate with keyboard input;
    \item \texttt{finish}: explicitly stop when the goal is satisfied.
\end{itemize}

The schema is deliberately application-neutral. A goal such as ``draft an
email'' is represented as visible UI actions over observed controls, not as a
backend mail API call.

\begin{table}[t]
\centering
\caption{Comparison of interaction strategies for AI computer interaction.}
\label{tab:comparison}
\setlength{\tabcolsep}{3pt}
\begin{tabular}{@{}p{0.20\linewidth}p{0.21\linewidth}p{0.23\linewidth}p{0.22\linewidth}@{}}
\toprule
\textbf{Approach} & \textbf{Strength} & \textbf{Weakness} & \textbf{LUMOS Position}\\
\midrule
Screenshot/VLM agents & General visual coverage & Token-heavy, coordinate uncertainty & Use only when semantic state is unavailable\\
RPA scripts & Reliable for fixed workflows & Brittle and task-specific & Avoid scripted task logic\\
Direct APIs & Efficient and precise & Not always available; bypasses visible UI & Prefer visible user-equivalent actions\\
Semantic blueprints & Compact and grounded & Depends on accessibility quality & Primary interaction substrate\\
\bottomrule
\end{tabular}
\end{table}

\subsection{Memory and Repair Layer}
The memory layer tracks recent actions, failures, and text already entered.
This matters because LLMs may repeatedly propose the same action or append text
that should replace previous content. LUMOS records when generated text was
typed into a target and can guide the model to either finish or use
\texttt{set\_text} for correction. It also stabilizes Windows Search handoffs:
when the model opens Search with a pending query such as ``outlook,'' LUMOS
ensures that the next step types that query into the Search overlay rather
than accidentally typing email content into a stale web field.

\subsection{Safety Layer}
LUMOS constrains the planner through an allowlist and confirmation policy.
Potentially risky operations, such as sending an email, deleting data, or
using hotkeys with destructive effects, require explicit confirmation. System
settings such as sound or display are accessed through visible settings UI
rather than hidden backend APIs. This preserves the principle that the agent
acts like a visible user, not an unrestricted system process.

\section{Prototype Implementation}
The prototype is implemented in Python and runs on Windows for native desktop
UI control, with browser support for web surfaces. Native observation uses
Windows UI Automation through Python automation libraries. Web observation
uses browser automation to extract page structure. The model client supports
local or compatible LLM backends. The runtime maintains a persistent browser
session for web tasks, a native blueprint for desktop windows, and a shared
ID map for action grounding.

For reproducible demos on slow local hardware, the repository also includes
opt-in deterministic scaffolds controlled by \texttt{LUMOS\_FAST\_PATHS=1},
including pre-launch, post-launch typing, finish-after-text, and search-overlay
helpers. These functions are disabled by default and are documented as ablation
knobs rather than hidden task bypasses. The architecture and case-study claims
in this paper refer to the default observe--LLM--act mode, where the model
selects task steps from the semantic blueprint and the runtime validates and
executes only visible primitives.

\begin{figure}[t]
\centering
\resizebox{\columnwidth}{!}{%
\begin{tikzpicture}[
    font=\scriptsize,
    card/.style={draw, rounded corners=2pt, thick, minimum width=4.55cm,
        align=center},
    arrow/.style={-{Stealth[length=2mm]}, thick},
    feedback/.style={-{Stealth[length=2mm]}, thick}
]
\node[card, fill=blue!6, minimum height=1.2cm] (screen) at (0,1.35)
{\textbf{Observed native window}\\\texttt{Untitled - Notepad}\\foreground window, 32 native elements};

\node[card, fill=green!8, minimum height=1.65cm] (bp) at (0,-0.35)
{\textbf{Semantic blueprint}\\
\begin{tabular}{@{}ll@{}}
\textbf{ID} & \texttt{A2}\\
\textbf{Role} & document/edit\\
\textbf{Action} & \texttt{type\_text}, \texttt{set\_text}\\
\textbf{State} & tracked text length
\end{tabular}};

\node[card, fill=orange!10, minimum height=0.85cm] (action) at (0,-2.05)
{\textbf{Grounded action}\\LLM targets \texttt{A2}; executor modifies visible Notepad};

\draw[arrow] (screen.south) -- node[right]{UIA extraction} (bp.north);
\draw[arrow] (bp.south) -- node[right]{ID map} (action.north);
\draw[feedback] (action.west) .. controls (-2.4,-2.05) and (-2.4,1.35) ..
    node[left, align=center]{visible\\feedback} (screen.west);
\end{tikzpicture}}
\caption{Notepad blueprint extraction from the debug traces. The visible
window is reduced to a compact semantic record: Notepad is foreground, the
native blueprint contains the text-entry target \texttt{A2}, and subsequent
typing or replacement actions are grounded to that identifier.}
\label{fig:blueprint}
\end{figure}
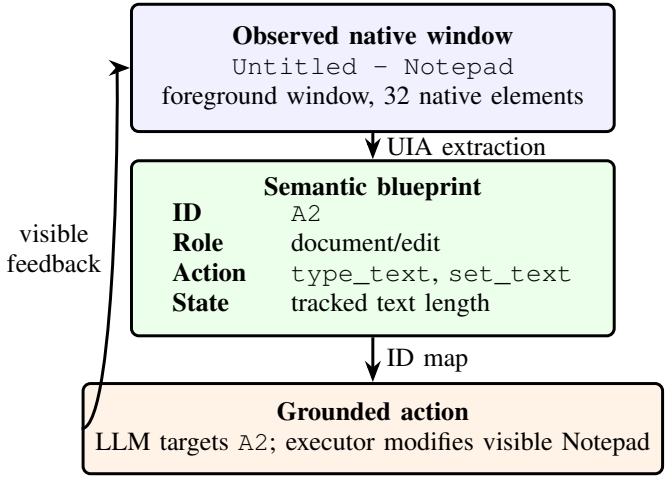

The implementation follows three rules:
\begin{enumerate}
    \item The LLM decides task strategy.
    \item The runtime exposes only universal perception and action primitives.
    \item The system re-observes after actions instead of assuming success.
\end{enumerate}

These rules distinguish the default LUMOS path from fixed workflow automation.
The runtime may repair invalid action syntax, prevent repeated app launches,
or convert append-style text entry into replacement when appropriate. Higher
level task sequencing remains a planner decision over the current blueprint,
and risky outcomes such as sending an email require explicit approval.

\section{Case Studies}
\subsection{Opening Notepad and Writing Generated Text}
The simplest demonstration is opening Notepad and asking the LLM to write
content. A literal instruction such as ``write a short essay about AI in
three paragraphs'' should not be typed verbatim. LUMOS therefore distinguishes
literal text-entry goals from generated-text goals. For generated-text goals,
the model must produce the content itself, while LUMOS tracks whether the
content has already been entered.

This case study surfaces an important design lesson: stopping is an action.
Without an explicit \texttt{finish} action, a model may continue revising,
retyping, or appending. LUMOS makes completion part of the action schema, so
the model can declare that the visible state satisfies the user goal.

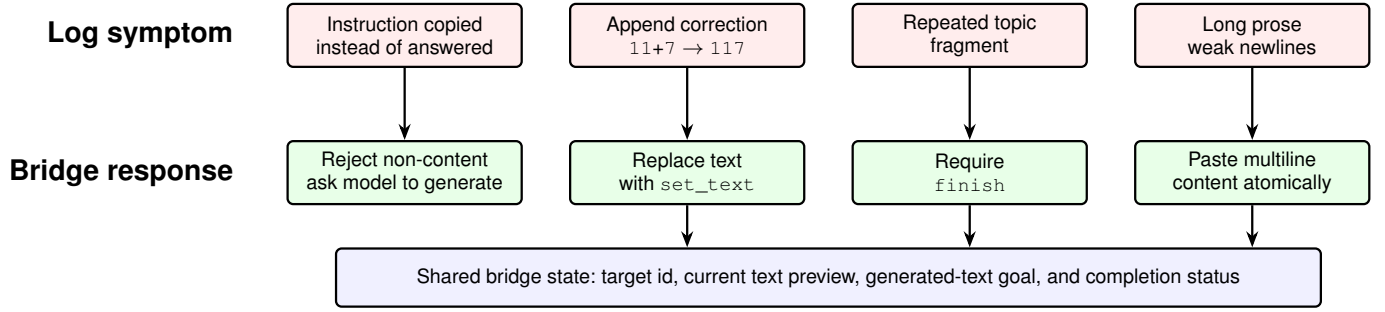
\begin{figure*}[t]
\centering
\resizebox{\textwidth}{!}{%
\begin{tikzpicture}[
    font=\sffamily\scriptsize,
    issue/.style={draw, rounded corners=2pt, thick, fill=red!7,
        minimum width=2.9cm, minimum height=0.78cm, align=center},
    repair/.style={draw, rounded corners=2pt, thick, fill=green!9,
        minimum width=2.9cm, minimum height=0.78cm, align=center},
    note/.style={draw, rounded corners=2pt, thick, fill=blue!6,
        minimum width=12.2cm, minimum height=0.72cm, align=center},
    arrow/.style={-{Stealth[length=2mm]}, thick}
]
\node[font=\sffamily\bfseries, anchor=east] at (-0.55,0.85) {Log symptom};
\node[font=\sffamily\bfseries, anchor=east] at (-0.55,-0.85) {Bridge response};

\node[issue] (i1) at (1.45,0.85) {Instruction copied\\instead of answered};
\node[issue] (i2) at (4.95,0.85) {Append correction\\\texttt{11}+\texttt{7} $\rightarrow$ \texttt{117}};
\node[issue] (i3) at (8.45,0.85) {Repeated topic\\fragment};
\node[issue] (i4) at (11.95,0.85) {Long prose\\weak newlines};

\node[repair] (r1) at (1.45,-0.85) {Reject non-content\\ask model to generate};
\node[repair] (r2) at (4.95,-0.85) {Replace text\\with \texttt{set\_text}};
\node[repair] (r3) at (8.45,-0.85) {Require\\\texttt{finish}};
\node[repair] (r4) at (11.95,-0.85) {Paste multiline\\content atomically};

\draw[arrow] (i1) -- (r1);
\draw[arrow] (i2) -- (r2);
\draw[arrow] (i3) -- (r3);
\draw[arrow] (i4) -- (r4);

\node[note] (mem) at (6.7,-2.15)
{Shared bridge state: target id, current text preview, generated-text goal, and completion status};
\coordinate (memr2) at (r2.south |- mem.north);
\coordinate (memr3) at (r3.south |- mem.north);
\coordinate (memr4) at (r4.south |- mem.north);
\draw[arrow] (r2.south) -- (memr2);
\draw[arrow] (r3.south) -- (memr3);
\draw[arrow] (r4.south) -- (memr4);
\end{tikzpicture}}
\caption{Diagnostic evidence extracted from the Notepad debug logs. Early
runs showed literal instruction copying, append-style corrections, repeated
fragments, and character-level typing problems for long prose. LUMOS repairs
these as semantic bridge failures: reject non-content, replace instead of
append, require explicit completion, and paste long or multiline content
through a stable text-entry path.}
\label{fig:notepad-diagnostics}
\end{figure*}

\subsection{Windows Search Handoff for an Outlook Query}
Desktop applications may not be directly available by executable name. In
such cases, a human would open Windows Search, type the app name, press Enter,
and observe the result. LUMOS follows the same visible workflow. When the
model emits:

\begingroup
\scriptsize
\begin{verbatim}
{"action":"open_windows_search",
 "text":"outlook"}
\end{verbatim}
\endgroup

the runtime opens Search and preserves the pending query. If the next
observation still shows Search, the system ensures that the query ``outlook''
is typed and submitted before the agent proceeds. This preserves LLM intent
while preventing stale web context from hijacking the next action; it is a
launch-handoff mechanism, not evidence that Outlook composition workflows have
been fully solved.

\section{Evaluation Plan}
A full evaluation should test whether semantic operating-system grounding
offers measurable advantages over screenshot/OCR-centric interaction. We
therefore propose four experiment families. First, vision versus
semantic grounding should compare a screenshot+OCR+LLM pipeline against a
LUMOS blueprint+LLM pipeline on identical tasks, measuring task success,
latency, token count, observation size, and number of recovery turns. Second,
blueprint compression should measure the size of raw screenshots,
OCR transcripts, vision-generated screen descriptions, and LUMOS blueprints
for the same UI states. Third, semantic pointer latency should compare
cursor-to-screenshot-crop interpretation against cursor-to-live-UIA query
for identifying the element under the pointer. Fourth, multi-step
desktop tasks should evaluate whether semantic blueprints help agents remain
grounded over long-horizon workflows in Notepad, Settings, browser search,
File Explorer, and mail clients after those applications are validated.

The current prototype is validated with regression tests for action schema
coercion, generated-text handling, text replacement, Windows Search handoff,
safety checks, and blueprint refresh behavior. These tests do not replace a
human-subject or benchmark evaluation, but they make the architectural claims
refereeable by showing that the system separates generated content from UI
actions, remembers visible text state, and repairs common model mistakes
without encoding application-specific scripts.

\begin{table}[t]
\centering
\caption{Proposed evaluation tasks and metrics.}
\label{tab:evaluation}
\begin{tabular}{p{0.28\linewidth}p{0.35\linewidth}p{0.27\linewidth}}
\toprule
Experiment & Comparison & Metrics\\
\midrule
Vision vs. semantic grounding & Screenshot+OCR+LLM vs. blueprint+LLM & Success, latency, tokens\\
Blueprint compression & Screenshot/OCR text vs. semantic blueprint & Observation size, token count\\
Semantic pointer latency & Cursor crop vs. UIA ElementFromPoint query & Identification latency, accuracy\\
Multi-step desktop tasks & Notepad, Settings, browser, File Explorer & Completion, steps, recovery\\
\bottomrule
\end{tabular}
\end{table}

\begin{figure}[t]
\centering
\resizebox{\columnwidth}{!}{%
\begin{tikzpicture}[
    font=\sffamily\tiny,
    bar/.style={draw, thick, fill=blue!35},
    repairbar/.style={draw, thick, fill=green!45},
    axis/.style={thick}
]
\draw[axis,->] (0,0) -- (0,4.9) node[above]{\textbf{count}};
\draw[axis,->] (0,0) -- (9.8,0);
\foreach \y/\lab in {0/0,1/5,2/10,3/15,4/20}
  \draw (-0.08,\y) -- (0.08,\y) node[left=2pt]{\textbf{\lab}};

\draw[bar] (0.55,0) rectangle (1.45,0.8);
\node[above, font=\sffamily\tiny\bfseries] at (1.0,0.8) {4};
\node[rotate=35, anchor=east] at (1.15,-0.18) {literal / fragment};

\draw[bar] (2.25,0) rectangle (3.15,1.0);
\node[above, font=\sffamily\tiny\bfseries] at (2.7,1.0) {5};
\node[rotate=35, anchor=east] at (2.85,-0.18) {append risks};

\draw[bar] (3.95,0) rectangle (4.85,4.4);
\node[above, font=\sffamily\tiny\bfseries] at (4.4,4.4) {22};
\node[rotate=35, anchor=east] at (4.55,-0.18) {repeat blocks};

\draw[repairbar] (5.65,0) rectangle (6.55,0.4);
\node[above, font=\sffamily\tiny\bfseries] at (6.1,0.4) {2};
\node[rotate=35, anchor=east] at (6.25,-0.18) {finish repairs};

\draw[repairbar] (7.35,0) rectangle (8.25,0.6);
\node[above, font=\sffamily\tiny\bfseries] at (7.8,0.6) {3};
\node[rotate=35, anchor=east] at (7.95,-0.18) {\texttt{set\_text} repairs};

\node[draw, rounded corners=2pt, fill=blue!12, align=left] at (4.85,5.25)
{Notepad debug traces: failure signals (blue) and bridge repairs (green)};
\end{tikzpicture}}
\caption{Diagnostic counts extracted from the pasted Notepad development
logs. They are not benchmark results; they summarize why the semantic bridge
needed literal-copy rejection, append-to-replace repair, repeat guards, and
explicit completion handling.}
\label{fig:results}
\end{figure}
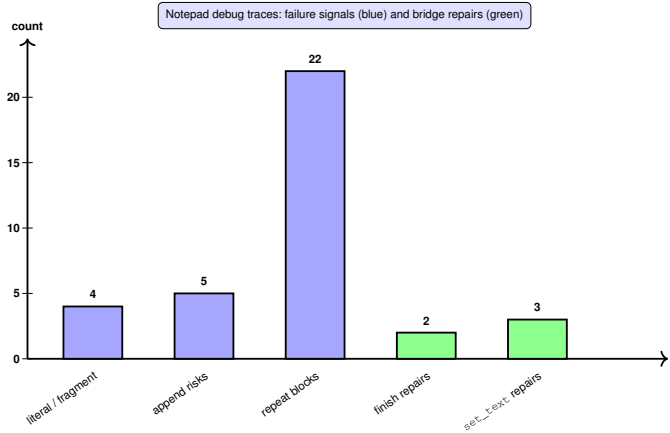

\section{Accessibility APIs as Cognitive Infrastructure}
Accessibility APIs were originally developed to make software usable by people
with diverse perceptual and motor abilities. Screen readers, switch devices,
voice-control tools, and UI testing frameworks depend on the fact that
interfaces can expose more than pixels. They expose roles, labels, values,
states, selection, focus, and interaction patterns. LUMOS repurposes this
infrastructure as a cognition layer for AI agents.

This reuse is technically important. A button's accessible name is a compact
semantic label. A text field's value provider exposes editable state. A control
pattern describes what operations are meaningful. A bounding rectangle connects
semantic identity to physical coordinates. Together, these properties form a
machine-readable contract between applications and external actors. In human
accessibility, the external actor may be a screen reader. In AI-native
computing, the external actor may be an LLM planner.

The implication is that accessibility infrastructure may become foundational
for future AI-native operating systems. Rather than treating accessibility as
an auxiliary compliance layer, operating systems could treat it as the core
semantic substrate through which agents understand and act. This does not
remove the need for human-centered design. It suggests that the same interface
can support both human perception and machine cognition when exposed through
parallel planes: a visual plane for people and a semantic plane for agents.

\section{Discussion}
\subsection{Why This is an Operating-System Problem}
The long-term implication of LUMOS is that AI agents need an operating-system
level interface designed for machine cognition. Current desktops provide a
human-facing interface and, separately, accessibility APIs for assistive
technologies. LUMOS treats those accessibility APIs as the first version of
an AI-native interaction plane.

Future operating systems could expose richer semantic state directly:
application intentions, available commands, reversible operations, security
boundaries, user approval requirements, and task progress. Rather than asking
an AI to infer everything from pixels, the OS could provide a trusted
machine-readable contract for what is visible, actionable, and safe.

\subsection{Why Not Only Screenshots?}
Screenshots remain valuable when applications do not expose useful semantics.
However, using screenshots as the default interface forces the model to solve
perception and action grounding simultaneously. Semantic blueprints separate
these concerns. The OS supplies structured state; the LLM supplies planning.
This separation is easier to test, cheaper to prompt, and more aligned with
security constraints.

\subsection{Why Not Only APIs?}
Direct APIs are efficient but often skip the interface the user can see. For
some tasks, bypassing the visible UI may violate user expectation or safety.
LUMOS favors visible UI actions because they are inspectable and reversible.
When a task says ``draft an email, do not send it,'' the model should open the
mail client, fill the draft, and stop before Send. It should not call a hidden
send API.

\section{Limitations}
LUMOS depends on the quality of exposed UI semantics. Some applications
provide incomplete accessibility trees, ambiguous names, duplicate controls,
or custom-rendered surfaces. Dynamic interfaces can change between observation
and action. LLMs may still choose incorrect actions, misunderstand task
completion, or require multiple recovery turns. Security remains a central
concern: an AI-controlled UI layer must prevent unintended submission,
deletion, credential exposure, or privilege escalation.

The prototype also does not claim human-level autonomy. Opening Notepad and
typing text is a small demonstration of the interaction model, not proof that
all desktop workflows are solved. The research value is in the architecture:
semantic extraction, grounded action IDs, constrained universal actions,
memory, safety, and explicit completion.

The present implementation is strongest on simple text-entry and launch tasks.
It can open Windows Search, carry a pending query such as an application name,
submit it, and re-observe the resulting UI. The remaining workflow is still
limited by the quality of the next blueprint and by the LLM's ability to choose
correct visible actions. This is why complex applications such as video editors,
mail clients, and custom-rendered professional tools are not yet demonstrated
as solved end-to-end. System-level operations also remain constrained: direct
backend volume and brightness APIs are intentionally removed, so LUMOS must open
the visible Settings UI and then identify and manipulate exposed sliders or
controls, with confirmation required for risky adjustment actions. These limits
point to the next engineering work: richer accessibility recovery, better
state verification, robust slider/control manipulation, and broader evaluation
on multi-application workflows.

\section{Conclusion}
Human-first operating systems are visually rich but not naturally optimized
for AI agents. LUMOS proposes a practical semantic interaction layer: use
existing accessibility and UI automation substrates to expose machine-readable
blueprints, let the LLM plan over those blueprints, ground actions through
live UI state and pointer semantics, and execute only constrained visible-UI
operations. This approach occupies a promising middle ground between
screenshot-heavy agents, brittle task scripts, and hidden backend APIs. More
broadly, LUMOS suggests that the next phase of AI-native computing may require
not only smarter models, but operating systems that expose explicit agent
interfaces alongside human user interfaces.

\section*{Code Availability}
The LUMOS repository is available at
\url{https://github.com/thotayogeswarreddy/Lumos.git}.

\bibliographystyle{IEEEtran}
\bibliography{references}

\end{document}